\documentclass[12pt]{article}
\catcode`\@=11
\@addtoreset{equation}{section}

\global\arraycolsep=2pt
\usepackage{mathrsfs,amsbsy,amssymb,latexsym,amsfonts,amsmath,cite}
\usepackage{graphicx,color}
\usepackage{physics}
\usepackage{mathtools}
\usepackage{ulem}

\newcommand{\pa}{\partial}

\newcommand{\no}{\nonumber}

\newcommand{\cL}{\mathcal L}
\newcommand{\cM}{\mathcal M}

\newcommand{\sfL}{\mathsf L}

\newcommand{\sfP}{\mathsf P}
\newcommand{\sfR}{\mathsf R}

\newcommand{\gP}{{\rm\bf P}}
\newcommand{\gJ}{{\rm\bf J}}

\newcommand{\gR}{{\rm \bf R}}
\newcommand{\bgg}{{\rm \bf g}}

\newcommand{\ha}{\hat{a}}
\newcommand{\hb}{\hat{b}}
\newcommand{\hc}{\hat{c}}

\newcommand{\cha}{\check{a}}
\newcommand{\chb}{\check{b}}
\newcommand{\chc}{\check{c}}
\newcommand{\chd}{\check{d}}

\newcommand{\llangle}{ \langle\!\langle}
\newcommand{\rrangle}{ \rangle\!\rangle}

\newcommand{\cc}[1]{\overline{#1}}

\newcommand{\Ad}{ {\rm Ad}}
\newcommand{\Str}{ {\rm Str}}

\allowdisplaybreaks

\begin{document}

\renewcommand{\thefootnote}{\fnsymbol{footnote}}

\begin{flushright}
KUNS-2817
\end{flushright}
\vspace*{0.5cm}

\begin{center}
{\Large \bf Yang-Baxter deformations of the AdS$_5\times$S$^5$ supercoset \\[7pt]
sigma model from 4D Chern-Simons theory}
\vspace*{2cm} \\
{\large  Osamu Fukushima$^{\sharp}$\footnote{E-mail:~osamu.f@gauge.scphys.kyoto-u.ac.jp},
Jun-ichi Sakamoto$^{\dagger}$\footnote{E-mail:~sakamoto@ntu.edu.tw},
and Kentaroh Yoshida$^{\sharp}$\footnote{E-mail:~kyoshida@gauge.scphys.kyoto-u.ac.jp}} 
\end{center}

\vspace*{0.4cm}

\begin{center}
$^{\sharp}${\it Department of Physics, Kyoto University, Kyoto 606-8502, Japan.}
\end{center}
\begin{center}
$^{\dagger}${\it Department of Physics and Center for Theoretical Sciences, National Taiwan University, Taipei 10617, Taiwan}
\end{center}

\vspace{1cm}

\begin{abstract}
We present homogeneous Yang-Baxter deformations of the AdS$_5\times$S$^5$ supercoset sigma model 
as boundary conditions of a 4D Chern-Simons theory. We first generalize the procedure for the 2D principal chiral model 
developed by Delduc et al [arXiv:1909.13824] so as to reproduce the 2D symmetric coset sigma model, 
and specify boundary conditions governing homogeneous Yang-Baxter deformations. 
Then the conditions are applicable for the AdS$_5\times$S$^5$ 
supercoset sigma model case as well. In addition, homogeneous bi-Yang-Baxter deformation is also discussed. 
\end{abstract}

\setcounter{footnote}{0}
\setcounter{page}{0}
\thispagestyle{empty}

\newpage

\tableofcontents

\renewcommand\thefootnote{\arabic{footnote}}

\section{Introduction}

A significant subject in mathematical physics is to establish a unified picture  
to describe integrable systems \cite{CWY1,CWY2}. By focusing upon 2D classical integrable systems 
including non-linear sigma models (NLSMs), such a nice way was originally proposed 
by Costello and Yamazaki \cite{CY} based on a 4D Chern-Simons (CS) theory with a meromorphic 1-form $\omega$\,.  
Notably, this 1-form $\omega$ is identified with a twist function characterizing the Poisson structure 
of the integrable system by Vicedo \cite{Gaudin}. Recently, this procedure has been elaborated 
by Delduc, Laxcroix, Magro and Vicedo \cite{DLMV} so as to describe systematic ways to perform integrable deformations 
of 2D principal chiral model (PCM) including the Yang-Baxter (YB) deformation 
\cite{Klimcik1,Klimcik2,DMV1,DMV2,Delduc:2014kha,KMY1,MY-CYBE} 
and the $\lambda$-deformation \cite{lambda1,lambda2}. 
For other recent works on this subject, see \cite{Schmidtt:2019otc,Bassi:2019aaf}. 

\medskip 

Our aim here is to generalize the preceding result on the PCM \cite{DLMV} 
to symmetric coset sigma models. By starting from 
a twist function in the rational description (with a slightly different parametrization of the spectral parameter), 
we specify a boundary condition associated with a symmetric coset. Then, the boundary condition is generalized 
so as to describe homogeneous YB deformations. It is straightforward to carry out the same analysis 
for the AdS$_5\times$S$^5$ supercoset sigma model. As a result, the homogeneous YB deformations of the AdS$_5\times$S$^5$ 
supercoset sigma model have been derived as specific boundary conditions of the 4D CS theory.  

\medskip 

This paper is organized as follows. Section 2 explains how to derive 2D NLSMs from 4D CS theory. 
In section 3, we derive 2D symmetric coset sigma models as boundary conditions of the 4D CS theory 
and then specify boundary conditions which describes homogeneous Yang-Baxter deformation. 
In section 4, the results obtained in section 3 are generalized to the AdS$_5\times$S$^5$ supercoset sigma model case. 
Section 5 is devoted to conclusion and discussion. Appendix A explains the computation concerned 
with a dressed $R$-operator in detail. In appendix B, we present homogeneous bi-Yang-Baxter deformed 
sigma models as boundary conditions of the 4D CS theory. 

\paragraph{NOTE:}
Just before submitting this manuscript to the arXiv, we have found an interesting work \cite{CS}. 
The content of \cite{CS} has some overlap with us on the integrability of the AdS$_5\times$S$^5$ 
superstring.

\section{2D NLSM from 4D CS theory} 

This section explains how to derive 2D NLSMs from a 4D CS theory by following \cite{CY,DLMV}.

\medskip

Let us begin with a 4D CS action\cite{CY}\footnote{For the notation and convention here, see \cite{FSY1}. },  
\begin{align}
S[A]=-\frac{i}{4\pi}\int_{\cM\times\mathbb{C}P^1} \omega\wedge CS(A)\,, 
\label{4dcs}
\end{align}
where $A$ is a $\mathfrak{g}^\mathbb{C}$-valued 1-form and $CS(A)$ is the CS 3-form defined as 
\begin{align}
CS(A)\equiv \left\langle A,dA+\frac{2}{3}A\wedge A\right\rangle\,. 
\end{align}
Then $\omega$ is a meromorphic 1-form defined as 
\begin{align}
\omega\equiv \varphi(z)dz \label{omega}
\end{align}
and $\varphi$ is a meromorphic function on $\mathbb{C}P^1$\,. This function is identified with a twist function 
characterizing the Poisson structure of the underlying integrable field theory \cite{Gaudin}.  

\medskip

Note that the $z$-component of $A$ can always be gauged away like  
\begin{align}
A=A_\sigma\, d\sigma+A_\tau\, d\tau+A_{\bar{z}}\, d\bar{z}\,, 
\end{align}
because $\varphi(z)$ depends only on $z$ and hence 
the action (\ref{4dcs}) has an extra gauge symmetry
\begin{align}
A\mapsto A+\chi\, dz\,. 
\label{extra gauge}
\end{align}

\medskip

The pole and zero structure of $\varphi$ will be important in the following discussion.  
The set of poles is denoted as $\mathfrak{p}$ and that of zeros is $\mathfrak{z}$\,.  
At each point of $\mathfrak{z}$, the 1-form $A$ 
cannot be regular because otherwise the action (\ref{4dcs}) is degenerate 
and hence the equations of motion at $\mathfrak{z}$ cannot be determined.

\medskip 

By taking a variation of the classical action (\ref{4dcs})\,, we obtain 
the bulk equation of motion  
\begin{align}
\omega\wedge F(A)=&0\,, \qquad F(A)\equiv dA+A\wedge A
\label{bulk eom}
\end{align}
and the boundary equation of motion 
\begin{align}
d\omega \wedge \langle A,\delta A\rangle=&0\,.\label{boundary eom}
\end{align}
Note that the boundary equation of motion (\ref{boundary eom}) has the support only on 
$\cM\times\mathfrak{p}\subset \cM\times \mathbb{C}P^1$\,, because 
\[
d \omega = \partial_{\bar{z}}\varphi(z)\, d\bar{z} \wedge dz 
\]
and only the pole of $\varphi$ can contribute as a distribution. 
The boundary conditions satisfying (\ref{boundary eom}) are crucial to describe integrable deformations \cite{CY,DLMV}.

\medskip

The bulk equation of motion (\ref{bulk eom}) can be expressed  in terms of the component fields:
\begin{align}
\partial_\sigma A_\tau -\partial_\tau A_\sigma+[A_\sigma,A_\tau]=&0\,,\\
\omega\,\left(\partial_{\bar{z}} A_\sigma -\partial_\sigma A_{\bar{z}}+[A_{\bar{z}},A_\sigma]
\right)=&0\,,\\
\omega\,\left(\partial_{\bar{z}} A_\tau -\partial_\tau A_{\bar{z}}+[A_{\bar{z}},A_\tau]
\right)=&0\,.
\end{align}
The factor $\omega$ is kept in order to cover the case $\partial_{\bar{z}}A_\sigma$ 
and $\partial_{\bar{z}}A_\tau$ are distributions on $\mathbb{C}P^1$ supported by $\mathfrak{z}$\,.

\medskip

It is also helpful to rewrite the boundary equation of motion (\ref{boundary eom})  into the form
\begin{align}
\sum_{x\in\mathfrak{p}}\sum_{p\geq0}\left(\operatorname{res}_x \xi_x^p \omega\right)\epsilon^{ij}\frac{1}{p!}\partial_{\xi_x}^p
\langle A_i,\delta A_j \rangle\big|_{\cM \times \{x\}}=0\,,\label{general boundary}
\end{align}
where $\epsilon^{ij}$ is the antisymmetric tensor.
Here the local holomorphic coordinates $\xi_x$ is defined as $\xi_x\equiv z-x$ 
for $x\in\mathfrak{p}\backslash\{\infty\}$ and $\xi_\infty\equiv1/z$ if $\mathfrak{p}$ 
includes the point at infinity. The relation (\ref{general boundary}) manifestly shows that the boundary equation 
of motion does not vanish only on $\cM\times\mathfrak{p}$\,.

\subsection*{Lax form}

By taking a formal gauge transformation
\begin{align}
A=-d\hat{g}\hat{g}^{-1}+\hat{g}\mathcal{L}\hat{g}^{-1} 
\label{L def}
\end{align}
with a smooth function $\hat{g}:\cM\times\mathbb{C}P^1
\rightarrow G^{\mathbb{C}}$, 
the following gauge is realized 
\begin{align}
\mathcal{L}_{\bar{z}}=0\,.  
\label{gauge fix}
\end{align}
Hence the 1-form $\mathcal{L}$ takes the form
\begin{align}
\mathcal{L}\equiv\mathcal{L}_{\sigma}d\sigma+\mathcal{L}_{\tau}d\tau\,,
\end{align}
and we call $\mathcal{L}$ the Lax form. This will be specified as a Lax pair for 2D theory later. 

\medskip 

In terms of the Lax form $\mathcal{L}$, the bulk equations of motion are expressed as
\begin{align}
\partial_\tau \mathcal{L}_\sigma -\partial_\sigma \mathcal{L}_\tau+&[\mathcal{L}_\tau,
\mathcal{L}_\sigma]=0\,,\\
\omega\wedge \partial_{\bar{z}}\mathcal{L}&=0\,.\label{L holomorphic}
\end{align}
It follows that $\mathcal{L}$ is a meromorphic 1-form with poles at the zeros of $\omega$\,,  
namely $\mathfrak{z}$ is regarded as the set of poles of $\mathcal{L}$\,. 

\subsection*{Reality condition}

It is natural to suppose some condition for the form of $\omega$ and its boundary condition on $A$ 
so as to ensure the reality of the 4D action (\ref{4dcs}) and the resulting action (\ref{2d action}) \cite{DLMV}.  

\medskip 

For a complex coordinate $z$, complex conjugation $z\mapsto \cc{z}$ defines an involution $\mu_{\mathrm{t}}:\mathbb{C}P^{1}\to\mathbb{C}P^{1}$\,.
Let $\tau:\mathfrak{g}^{\mathbb{C}}\to\mathfrak{g}^{\mathbb{C}}$ be an anti-linear involution. 
Then the set of the fixed point under $\tau$ is a real Lie subalgebra $\mathfrak{g}$ of $\mathfrak{g}^{\mathbb{C}}$.
The anti-linear involution $\tau$ satisfies
\begin{align}
\cc{\langle B,C \rangle} = \langle \tau B, \tau C \rangle\,, \qquad ^{\forall}B,C\in \mathfrak{g}^{\mathbb{C}}\,.
\end{align}
The associated operation to the Lie group $G$ is denoted by $\tilde{\tau}: G^{\mathbb{C}}\to G^{\mathbb{C}}$\,.

\medskip

Introducing the involutions, one can see that the reality of the action (\ref{2d action}) is ensured by the conditions
\begin{align}
\cc{\omega}=&\mu_{\mathrm{t}}^{*}\omega\,,\label{reality omega}  \\
\tau A=&\mu_{\mathrm{t}}^{*}A\,.\label{reality A}  
\end{align}
Recalling the relation (\ref{L def}), we suppose that 
\begin{align}
 \tilde{\tau} \hat{g}=\mu_{\mathrm{t}}^{*}\hat{g}\,,\qquad & \tau \mathcal{L}=\mu_{\mathrm{t}}^{*}\mathcal{L}\,,\label{reality L}
\end{align}
so as to satisfy (\ref{reality A}).

\subsection*{From 4D to 2D via the archipelago conditions}

When $\hat{g}$ satisfies the archipelago conditions \cite{DLMV}, the 4D action (\ref{4dcs}) is reduced to 
a 2D action with the WZ term by performing an integral 
over $\mathbb{C}P^1$ as follows: 
\begin{align}
S\left[\left\{g_{x}\right\}_{x \in \mathfrak{p}}\right]=\frac{1}{2} \sum_{x \in \mathfrak{p}} \int_{\cM}\left\langle\operatorname{res}_{x} (\varphi \,\mathcal{L}), g_{x}^{-1} d g_{x}\right\rangle
+\frac{1}{2} \sum_{x \in \mathfrak{p}}\left(\operatorname{res}_{x} \omega\right) \int_{\cM\times[0,R_x]} I_{\mathrm{WZ}}\left[g_{x}\right]\,. 
\label{2d action}
\end{align}
Here $R_x$ is the radius of the open disk $U_x$\,. 

\medskip

The action (\ref{2d action}) is invariant under a gauge transformation 
\begin{align}
g_x\mapsto g_x h\,,\qquad \mathcal{L}\mapsto h^{-1}\mathcal{L}h+h^{-1}dh\,, 
\label{2d gauge}
\end{align}
with a local function $h:\cM\rightarrow G^{\mathbb{C}}$\,. 
This gauge symmetry can be seen as the remnant after taking the gauge (\ref{gauge fix})\,. 
Note here that we have not imposed the reality condition by following \cite{FSY1}, 
in comparison to \cite{DLMV}. The reality condition will be introduced later 
when fixing a boundary condition of $\hat{g}$\,.

\section{YB deformations of the symmetric coset sigma model}

In this section, we will reproduce the action of a symmetric coset sigma model and 
homogeneous Yang-Baxter deformations of it from the $4$D CS theory (\ref{4dcs}) 
by generalizing the work \cite{DLMV,FSY1}.  
The symmetric coset case has been discussed in \cite{CY} in a slightly different way. 

\subsubsection*{Symmetric coset}

Let $G$ and $H$ be a Lie group and its subgroup, and
the Lie algebras for $G$ and $H$ are denoted as $\mathfrak{g}$ and $\mathfrak{h}$\,, respectively.
We assume that the Lie algebra $\mathfrak{g}$ enjoys a $\mathbb{Z}_2$-grading, namely,  
$\mathfrak{g}$ is decomposed like $\mathfrak{g} = \mathfrak{h} \oplus \mathfrak{m}$ as the vector space 
and the following relations are satisifed 
\begin{align}
    [\mathfrak{h},\mathfrak{h}]\subset\mathfrak{h}\,,\qquad [\mathfrak{h}, \mathfrak{m}]\subset \mathfrak{m}\,,\qquad
    [\mathfrak{m},\mathfrak{m}]\subset \mathfrak{h}\,.
    \label{Z2-symm}
\end{align}

\subsubsection*{Twist function} 
The twist function for a symmetric coset sigma model is given by\footnote{The twist function (\ref{eq:twist-new}) 
is the same as the one for PCM, and they are related by a transformation
\begin{align}
    z=\frac{1+z'}{1-z'}\,, 
\end{align}
where $z'$ is the spectral parameter for PCM.
} 
\begin{align}
    \omega=\varphi_{c}(z)\,dz=\frac{16K z}{(z-1)^2(z+1)^{2}}dz\,,
    \label{eq:twist-new}
\end{align}
where we have followed the notation in \cite{DLMV-bi}.
The meromorphic 1-form $\omega$ indeed satisfies the reality condition (\ref{reality omega}).
The poles and zeros of $\varphi_{c}(z)$ are listed as 
\begin{align}
    \mathfrak{p}=\{\pm 1\}\,,\qquad \mathfrak{z}=\{0, \infty\}\,,
\end{align}
where these poles are double poles, and each zero is a single zero. 
As we will see later, the twist function (\ref{eq:twist-new}) is applicable not only to symmetric cosets, 
but also to homogeneous YB deformed sigma models.

\subsubsection*{Boundary condition}

In order to specify a 2D integrable model, we need to choose a solution 
to the boundary equations of motion, 
\begin{align}
   \epsilon^{ij} \llangle(A_{i},\partial_{\xi_{p}} A_{i}),\delta(A_{j}, \partial_{\xi_{p}} A_{j})\rrangle_{p}=0\,,\qquad p\in\mathfrak{p}\,.
   \label{eq:beom}
\end{align}
Here the double bracket is defined as
\begin{align}
    \llangle (x,y), (x',y')\rrangle_{p}&\equiv(\text{res}_{p}\,\omega)\langle x, x'\rangle+(\text{res}_{p}\,\xi_{p}\omega)\left(\langle x, y'\rangle+\langle x', y\right\rangle)\no\\
    &=4p\,K\left(\langle x, y'\rangle+\langle x', y\rangle\right)\,.
    \label{def:inner-double}
\end{align}
The boundary equations of motion (\ref{eq:beom}) take the same form as in the PCM case.

\medskip

In the following, we will consider two classes of solutions. 
Note that $A|_{z=\pm1}$ and $\pa_{z}A|_{z=\pm1}$ take values in the real Lie algebra $\mathfrak{g}$, supposing the reality conditions (\ref{reality omega}) and (\ref{reality A}) and the points $z=\pm1$ are fixed points of the involution $\mu_{\mathrm{t}}$\,.

\medskip

The first class is
\begin{align}
\text{i)}\qquad (A|_{z=1},\partial_{z} A|_{z=1})\in \{0\}\ltimes\mathfrak{g}_{\rm ab}\,,
\qquad (A|_{z=-1},\partial_{z} A|_{z=-1})\in \{0\}\ltimes\mathfrak{g}_{\rm ab}\,,
\label{eq:A-bsol}
\end{align}
where $\{0\}\ltimes\mathfrak{g}_{\rm ab}$ is an abelian copy of $\mathfrak{g}$ defined as
\begin{align}
    \{0\}\ltimes\mathfrak{g}_{\rm ab}\equiv \left\{\left(0,x\right)\,|\,x\in\mathfrak{g}\right\}\,.
    \label{semi-ab}
\end{align}
This configuration obviously solves the boundary equations of motion and lead to a symmetric coset sigma model 
as we will see later.

\medskip

The second class is
\begin{align}
\text{ii)}\qquad (A|_{z=1},\partial_{z} A|_{z=1})\in \mathfrak{g}_{R}\,,\qquad (A|_{z=-1},\partial_{z} A|_{z=-1})\in \mathfrak{g}_{\tilde{R}}\,,
\label{eq:A-bsol-hYB}
\end{align}
where $\mathfrak{g}_{R}$ and $\mathfrak{g}_{\tilde{R}}$ are defined as
\begin{align}
    \mathfrak{g}_{R}\equiv\left\{\left( 2\eta R(x),x\right)\,|\,x\in\mathfrak{g}\right\}\,,\qquad 
    \mathfrak{g}_{\tilde{R}}\equiv \{(- 2\eta \tilde{R}(x),x)\,|\,x\in\mathfrak{g}\}\,.
    \label{hYB-DD}
\end{align}
Here the linear $R$-operator $R:\mathfrak{g}\to \mathfrak{g}$ satisfies 
the homogeneous classical Yang-Baxter equation (hCYBE), 
\begin{align}
 [R(x),R(y)]-R([R(x),y]+[x,R(y)])=0\,,\qquad x\,,y \in \mathfrak{g}\,.
    \label{eq:hCYBE}
\end{align}
The other $R$-operator $\tilde{R}:\mathfrak{g}\to \mathfrak{g}$ is defined as
\begin{align}
  \tilde{R}\equiv f\circ R \circ f  \,,
  \label{eq:tR}
\end{align}
where $f:\mathfrak{g}\to \mathfrak{g}$ is a $\mathbb{Z}_2$-grading automorphism 
of $\mathfrak{g}$\,. An explicit represetation will be given in (\ref{eq:sym-map}).
For $f$ in (\ref{eq:sym-map})\,, 
we can show that $\tilde{R}$ also solves the hCYBE (\ref{eq:hCYBE}) if the $R$-operator $R$ is a solution to the equation (\ref{eq:hCYBE}).
Furthermore, thanks to the hCYBE (\ref{eq:hCYBE}), we can check that the the second configuration in (\ref{eq:A-bsol-hYB}) 
solves the boundary equations of motion (\ref{eq:beom}).
The choice of the boundary conditions (\ref{eq:A-bsol-hYB}) is motivated by the one of homogeneous bi-YB deformations 
(For the details, see appendix \ref{sec:bi-YB}).
Note that the first and the second solutions are related by a $\beta$-transformation at the Lie algebra level 
(For the details, see appendix A of \cite{FSY1}).

\subsubsection*{Lax form}

Before deriving sigma model actions, we shall summarize our notation used in the following.
We will take $\hat{g}$ at each pole of the twist function (\ref{eq:twist-new}) as
\begin{align}
    \hat{g}(\tau,\sigma,z)|_{z=1}&=g(\tau,\sigma)\,,\qquad \hat{g}(\tau,\sigma,z)|_{z=-1}=\tilde{g}(\tau,\sigma)\,,
\end{align}
where $g\,, \tilde{g} \in G$\,. 
Here $g$ and $\tilde{g}$ take values in $G$ (not $G^{\mathbb{C}}$) due to the reality condition (\ref{reality L})\,.
%The reality condition has been implicitly imposed at this moment.
The associated left-invariant currents are defined as
\begin{align}
    j& \equiv g^{-1}dg\,, \qquad \tilde{j} \equiv \tilde{g}^{-1}d\tilde{g}\,.
\end{align}
Then, the relation between the gauge field and the Lax pair at each pole becomes    
\begin{align}
    A|_{z=1}=-dg g^{-1}+{\rm Ad}_{g}\cL|_{z=1}\,,\qquad
    A|_{z=-1}=-d\tilde{g}\tilde{g}^{-1}+{\rm Ad}_{\tilde{g}}\cL|_{z=-1}\,.
    \label{eq:A-Lax-symm}
\end{align}
From the zeros of the twist function (\ref{eq:twist-new}),
we suppose an ansatz for the Lax pair as
\begin{align}
    \cL=(U_{+}+z\,V_{+})d\sigma^++(U_{-}+z^{-1}\,V_{-})d\sigma^{-}\,,\label{eq:Lax-ansatz}
\end{align}
where $U_{\pm}\,, V_{\pm}\in \mathfrak{g}$ are undetermined functions of $\sigma\,, \tau$\,, and 
the light-cone coordinates are defined as
\begin{align}
\sigma^{\pm} \equiv \frac{1}{2}\left(\tau\pm\sigma\right)\,.
\end{align} 
As we will see, the ansatz (\ref{eq:Lax-ansatz}) of the Lax pair works well for the two classes of 
boundary conditions.

\medskip

\subsection*{i) symmetric coset sigma model}

Let us first see the class i) that describes a symmetric coset sigma model.

\medskip

Under the boundary condition (\ref{eq:A-bsol}), the relations in (\ref{eq:A-Lax-symm}) are rewritten as 
\begin{align}
    j_{\pm}&=U_{\pm}+V_{\pm}\,,\qquad
    \tilde{j}_{\pm}=U_{\pm}-V_{\pm}\,.   
\end{align}
By solving these equations with respect to $U_{\pm}$ and $V_{\pm}$\,, we obtain
\begin{align}
    U_{\pm}&=\frac{ j_{\pm}+\tilde{j}_{\pm}}{2}\,,\qquad
    V_{\pm}=\frac{ j_{\pm}-\tilde{j}_{\pm}}{2}\,. 
\end{align}
As a result, the Lax pair is expressed as 
\begin{align}
    \cL_{\pm}&=  \frac{j_{\pm}+\tilde{j}_{\pm}}{2}+z{}^{\pm1}\,\frac{ j_{\pm}-\tilde{j}_{\pm}}{2}\,.\label{eq:Lax}
\end{align}
Then, the residues of $\varphi_{c}\, \cL$ at $z=\pm 1$ are evaluated as
\begin{align}
\begin{split}
    \text{res}_{z=1}(\varphi_{c}\, \cL)&=4K(V_{+}d\sigma^{+}-V_{-}d\sigma^{-})\,,\\
        \text{res}_{z=-1}(\varphi_{c}\, \cL)&=-4K(V_{+}d\sigma^{+}-V_{-}d\sigma^{-})\,.
        \label{eq:phiL-res}
        \end{split}
\end{align}
By substituting (\ref{eq:phiL-res}) into (\ref{2d action}), the $2$D action is given by 
\begin{align}
    S[g,\tilde{g}]=K\int_{\cM}\left\langle j_{+}-\tilde{j}_+,j_{-}-\tilde{j}_{-}\right\rangle d\sigma \wedge d\tau\,.
    \label{eq:sym-action-lr}
\end{align}
If $\tilde{g}$ is independent of $g$, then by using the gauge symmetry of the 4D CS theory, 
we can rewrite the $2$D action (\ref{eq:sym-action-lr}) to that of PCM with Lie group $G$\cite{CY,DLMV}.

\medskip

Here we would like to impose a relation between $j$ and $\tilde{j}$\,. 
Note that the resulting action (\ref{eq:sym-action-lr}) is invariant under the exchange of $j$ and $\tilde{j}$\,. 
This invariance should be respected in a relation $\tilde{j}=f(j)$ and hence the automorphism $f:\mathfrak{g}\to \mathfrak{g}$ 
should satisfy the following conditions: 
\begin{align}
    f([x,y])=[f(x),f(y)]\,,\qquad f\circ f(x)=x\,,\qquad x\,,y \in \mathfrak{g}\,.
\end{align}
In order to obtain the known result, we will take $f$ satisfying the following relations: 
\begin{align}
f(\gP_{\cha})=-\gP_{\cha}\,,\qquad f(\gJ_{\ha})=\gJ_{\ha}\,. 
\label{eq:sym-map}
\end{align}
Here we have introduced the generators of the decomposed vector space 
$\mathfrak{g}=\mathfrak{h}\oplus \mathfrak{m}$ as
\begin{align}
   \mathfrak{h} = \langle \gJ_{\ha} \rangle\,, 
\qquad 
\mathfrak{m} = \langle \gP_{\cha} \rangle \,,
\end{align}
where $\ha=1,\dots, {\rm dim}\,\mathfrak{h}$ and $\cha=1,\dots, {\rm dim}\,\mathfrak{m}$\,.

\medskip

By employing the automorphism (\ref{eq:sym-map}), $\tilde{j}$ is evaluated as
\begin{align}
    \tilde{j}=f(j)=f\left(P_{(0)}(j)+P_{(2)}(j)\right)=P_{(0)}(j)-P_{(2)}(j)\,,
    \label{tg-conf}
\end{align}
where the projection operators $P_{(0)}$ and $P_{(2)}$ are defined as, respectively,   
\begin{align} 
P_{(0)}~:~\mathfrak{g} \rightarrow \mathfrak{h}\,, \qquad 
P_{(2)}~:~\mathfrak{g} \rightarrow \mathfrak{m}\,.
\end{align}
Then, by using the expression of $\tilde{j}$ in (\ref{tg-conf}), the 2D action can be further rewritten as 
\begin{align}
    S[g]=4K\int_{\cM}\left\langle j_{+},P_{(2)}(j_-)\right\rangle d\sigma \wedge d\tau\,,
\end{align}
and the Lax pair (\ref{eq:Lax}) becomes
\begin{align}
    \cL_{\pm}=P_{(0)}(j_{\pm})+z^{\pm1}\,P_{(2)}(j_{\pm})\,.
\end{align}
These are the standard expressions of the classical action and the associated Lax pair 
for a symmetric coset sigma model.

\subsection*{ii) homogeneous YB deformations}

The next one we will discuss is the class ii) in (\ref{eq:A-bsol-hYB}) 
that describes homogeneous YB deformations 
of a symmetric coset sigma model. 

\medskip

The condition gives a constraint on the gauge field $A$ at each pole of the twist function,
\begin{align}
    A|_{z=1}=2\eta\,R(\partial_{z}A|_{z=1})\,,\qquad 
      A|_{z=-1}=-2\eta\,\tilde{R}(\partial_{z}A|_{z=-1})\,.   
      \label{eq:A-constraint-hYB}
\end{align}
We again suppose the same ansatz (\ref{eq:Lax-ansatz}) for the Lax pair.
Then, the constraints in (\ref{eq:A-constraint-hYB}) lead to
\begin{align}
    j_{\pm}&=U_{\pm}+(1\mp2\eta R_{g})(V_{\pm})\,,\qquad
    \tilde{j}_{\pm}=U_{\pm}-(1\mp2\eta \tilde{R}_{\tilde{g}})(V_{\pm})\,,   
\end{align}
where we defined $R_{g}\equiv {\rm Ad}_{g^{-1}}\circ  R\circ {\rm Ad}_{g}$\,.
By solving these equations with respect to $U_{\pm}$ and $V_{\pm}$\,, we obtain 
\begin{align}
    U_{\pm}&=\frac{j_{\pm}+\tilde{j}_{\pm}}{2}\pm\eta(R_{g}-\tilde{R}_{\tilde{g}})(V_{\pm})\,,\qquad
    V_{\pm}=\frac{1}{1\mp\eta R_{g}\mp \eta \tilde{R}_{\tilde{g}}}\left(\frac{  j_{\pm}-\tilde{j}_{\pm}}{2}\right)\,.\label{eq:V-hYB}
\end{align} 
The residues of $\varphi_{c}\,\cL$ at $z=\pm 1$ take the same forms as (\ref{eq:phiL-res}), but $V_{\pm}$ are given by (\ref{eq:V-hYB}).
Thus the $2$D action is given by
\begin{align}
    S[g,\tilde{g}]=4K\int_{\cM}\left\langle \frac{  j_{+}-\tilde{j}_{+}}{2},\frac{1}{1+\eta R_{g}+\eta \tilde{R}_{\tilde{g}}}
    \left(\frac{  j_{-}-\tilde{j}_{-}}{2}\right)\right\rangle d\sigma \wedge d\tau\,.
    \label{eq:hYB-action-lr}
\end{align}
Note that in the present case, the resulting action is invariant under the exchange of $g$ and $\tilde{g}$, not $j$ and $\tilde{j}$.

\medskip

The exchange symmetry of the action (\ref{eq:hYB-action-lr}) at the level of group element leads to 
a slight change in the previous case:
for group elements, we impose an additional condition
\begin{align}
\tilde{g}=F(g)\,,
\label{tg=Fg}
\end{align}
where an automorphism $F:G\to G$ has the $\mathbb{Z}_2$-grading property $F\circ F(g)=g$\,.
To specify an explicit representation of $F$\,, let us take a parameterization of an element $g\in G$ as
\begin{align}
    g=\exp(X^{\cha}\gP_{\cha}+X^{\ha}\gJ_{\ha})\,,
\end{align}
where $X^{\cha}$ and $X^{\ha}$ are functions of $\tau$ and $\sigma$\,.
Then, in a neighborhood of the identity, $F(g)$ can be written by using the automorphism $f:\mathfrak{g}\to \mathfrak{g}$ as follows,
\begin{align}
    F(g)\equiv\exp(X^{\cha}f(\gP_{\cha})+X^{\ha}f(\gJ_{\ha}))\,.\label{Fg-coset}
\end{align}
or equivalently,
\begin{align}
    F(g)=\exp(-X^{\cha}\gP_{\cha}+X^{\ha}\gJ_{\ha})\,.\label{Aut-ad}
\end{align}

\medskip

Now let us rewrite the 2D action (\ref{eq:hYB-action-lr}) by requiring (\ref{tg=Fg}). 
As shown in appendix \ref{sec:proofRgRtg}, we can show that the dressed $R$-operators $R_g$ 
and $\tilde{R}_{\tilde{g}}$ satisfy the following relation:
\begin{align}
    ( R_{g}+ \tilde{R}_{\tilde{g}})\circ P_{(2)}(x)
    &=2P_{(2)}\circ R_{g}\circ  P_{(2)}(x)\,.
    \label{eq:RgRtg}
\end{align}
The relation (\ref{eq:RgRtg}) indicates 
\begin{align}
    V_{\pm}=P_{(2)}\left(\frac{1}{1\mp2\eta R_{g}\circ P_{(2)}}j_{\pm}\right)\,.
    \label{eq:V-relation}
\end{align}
Furthermore, by using (\ref{eq:V-relation}), $U_{\pm}$ can be rewritten as
\begin{align}
    U_{\pm}= j_{\pm}-(1\mp\eta R_{g})(V_{\pm})
    =P_{(0)}\left(\frac{1}{1\mp2\eta R_{g}\circ P_{(2)}}j_{\pm}\right)\,.
\end{align}
As a result, we obtain the 2D action
\begin{align}
    S[g]=4K\int_{\cM}\left\langle j_{-},P_{(2)}\left(\frac{1}{1-2\eta R_{g}\circ P_{(2)}}j_{+}\right)\right\rangle d\sigma \wedge d\tau\,,
    \label{eq:2Daction-hYB-coset}
\end{align}
and the Lax pair
\begin{align}
    \cL_{\pm}=P_{(0)}\left(\frac{1}{1\mp2\eta R_{g}\circ P_{(2)}}j_{\pm}\right)+z^{\pm1}\,P_{(2)}\left(\frac{1}{1\mp2\eta R_{g}\circ P_{(2)}}j_{\pm}\right)\,.\label{Lax-hYB-coset}
\end{align}
These are the standard expressions of the classical action and the Lax pair for a homogeneous YB deformed 
symmetric coset sigma model \cite{MY-CYBE}.

\section{YB deformations of the $\mathrm{AdS}_5\times \mathrm{S}^5$ supercoset sigma model}

In this section, we shall reproduce the Green-Schwarz (GS) action of 
the $\mathrm{AdS}_5\times \mathrm{S}^5$ supercoset sigma model \cite{Metsaev:1998it} 
and homogeneous YB deformations of it \cite{KMY1} from the 4D CS theory.

\subsubsection*{Supercoset}

The action of the $\mathrm{AdS}_5\times \mathrm{S}^5$ superstring in the GS formalism \cite{Metsaev:1998it} 
is based on the following supercoset
\begin{align}
\frac{PSU(2,2|4)}{SO(1,4)\times SO(5)}\,.
\end{align}
The gauge field $A$ in the 4D CS action (\ref{4dcs}) takes a value in $\mathfrak{g}=\mathfrak{su}(2,2|4)$\,.
Usually, $\mathfrak{su}(2,2|4)$ is represented by using $8\times8$ supermatrices satisfying the supertraceless and the relaity conditions. 
Then the bracket $\langle \cdot,\cdot \rangle$ in the 4D action (\ref{4dcs}) is replaced by the supertrace $\Str$\,.

\subsubsection*{Twist function}

The Poisson structure of the $\mathrm{AdS}_5\times \mathrm{S}^5$ superstring has been considered 
in \cite{Itoyama:2006cg,Vicedo:2009sn}, and the twist function of the $\mathrm{AdS}_5\times \mathrm{S}^5$ supercoset sigma model is given by\footnote{$\varphi_{\rm str}(z)$ is slightly different from $\phi_{\rm string}(z)$ in (2.10) of \cite{Delduc:2014kha}. These are related via $\varphi_{\rm str}(z)=\frac{1}{z}\,\phi_{\rm string}(z)$\,. }
\begin{align}
 \omega=\varphi_{\rm str}(z)\,dz=\frac{4z^3}{(z^4-1)^2}\,d z\,.
    \label{eq:twist-super}
\end{align}
%\red{In this section, we define the involution $\mu_{\rm t}:\mathbb{C}\to \mathbb{C}$ as
%\begin{align}
%\mu_{\rm t}: z\mapsto z'=\frac{1}{\bar{z}}\,.
%\end{align}
%The meromorphic 1-form (\ref{eq:twist-super}) satisfies the reality condition (\ref{reality omega}) under $\mu_{\rm t}$:
%\begin{align}
%\cc{\omega}=\frac{-4 i \bar{z}^3}{(\bar{z}^4-1)^2}d\bar{z}
%=\mu_{\rm t}^{*}\left(\frac{- 4 i z^{-3}}{(z^{-4}-1)^2}d\bigg(\frac{1}{z}\bigg)\right)
%=\mu_{\rm t}^{*}\omega\,.
%\end{align}
%}
Here $\omega$ is invariant under the involution $\mu_{\mathrm{t}}$ since $\varphi_{\mathrm{str}}$ satisfies
\begin{align}
\overline{\varphi_{\rm str}(z)}=\varphi_{\rm str}(\bar{z})\,.
\end{align}
The poles and zeros of the twist function (\ref{eq:twist-super}) are listed as 
\begin{align}
    \mathfrak{p}=\{+1\,,-1\,,+i\,,-i\}\,,\qquad \mathfrak{z}=\{0, \infty\}\,,
\end{align}
where the poles are double poles and the zeros are triple zeros.

\subsubsection*{Boundary condition}

The associated boundary equations of motion are
\begin{align}
   \epsilon^{ij} \llangle(A_{i},\partial_{\xi_{p}} A_{i}),\delta(A_{j}, \partial_{\xi_{p}} A_{j})\rrangle_{p}=0\,,\qquad p\in\mathfrak{p}\,,
   \label{eq:beom-sol}
\end{align}
where the double bracket is defined as
\begin{align}
    \llangle (x,y), (x',y')\rrangle_{p}&\equiv(\text{res}_{p}\,\omega)\,\Str( x \cdot  x')+(\text{res}_{p}\,\xi_{p}\omega)\left(\Str( x\cdot y')+\Str( x'\cdot y\right))\no\\
    &=\frac{p}{4}\left(\Str( x\cdot y')+\Str( x'\cdot y)\right)\,.
    \label{def:inner-double-super}
\end{align}

\medskip

As in the symmetric coset case, one may consider two classes of solutions 
to the boundary equations of motion (\ref{eq:beom-sol}). 
%%%%%%%%%%%%%%%%%
Now by considering 
\begin{align}\begin{split}
\tau\big((\pa_{z}A)|_{z=p}\big)=&\big( \tau(\pa_{z} A)\big)|_{z=p}
=\big(\pa_{\bar{z}}(\tau A) \big)|_{z=p}
=\big(\mu_{\mathrm{t}}^{*}(\pa_{z} A)\big)|_{z=p}\\
=&(\pa_{z} A)|_{z=\cc{p}}\,,
\end{split}\end{align}
the reality condition (\ref{reality A}) leads to 
\begin{align}
\big(\tau(A|_{z=p}),\tau(\partial_{z} A|_{z=p})\big)=\big(A|_{z=\cc{p}},\left(\partial_{z} A\right)|_{z=\cc{p}}\big)\,.
\label{eq:reality-A}
\end{align}

Due to the reality condition (\ref{eq:reality-A})\,, the relation
\begin{align}
(\pa_{z}A)|_{z=+ i}=(\pa_{z}A)|_{z=- i}\quad \left(\in \mathfrak{su}(2,2|4)^{\mathbb{C}}\right)
\end{align}
holds for the following two boundary conditions.

%%%%%%%%%%%%%%%%%
%\red{Now taking into account
%\begin{align}\begin{split}
%\tau\big((\pa_{z}A)|_{z=p}\big)=&\big( \tau(\pa_{z} A)\big)|_{z=p}
%=\big(\pa_{\bar{z}}(\tau A) \big)|_{z=p}
%=\big(\mu_{\mathrm{t}}^{*}(-z^2\pa_{z} A)\big)|_{z=p}\\
%=&(-z^2\pa_{z} A)|_{z=1/\cc{p}}\,,
%\end{split}\end{align}
%the reality condition (\ref{reality A}) implies
%\begin{align}
%\big(\tau(A|_{z=p}),\tau(\partial_{z} A|_{z=p})\big)=\big(A|_{z=1/\cc{p}},\left(-z^2\partial_{z} A\right)|_{z=1/\cc{p}}\big)\,.
%\end{align}}
%%%%%%%%%%%%%%%%%%

\medskip

For the $\mathrm{AdS}_5\times \mathrm{S}^5$ supercoset sigma model, we take the following solution:
\begin{align}
 \text{i)}\qquad   (A|_{z=p},\partial_{z} A|_{z=p})&\in \{0\}\ltimes\mathfrak{su}(2,2|4)_{p,\rm ab}\qquad (p\in \mathfrak{p})
\,,
\label{eq:A-bsol-super}
\end{align}
where $\mathfrak{su}(2,2|4)_{p,\rm ab}$ is an abelian copy of $\mathfrak{su}(2,2|4)$ and $\mathfrak{su}(2,2|4)^\mathbb{C}$\,:
\begin{align}
\mathfrak{su}(2,2|4)_{p, \rm ab}\equiv
\left\{\begin{array}{ll}
\mathfrak{su}(2,2,|4) \qquad& \mbox{for }p=\pm1 \\
\mathfrak{su}(2,2,|4)^\mathbb{C} & \mbox{for }p=\pm i\,.
\end{array}\right.
\end{align}
The second choice for a homogeneous YB deformed $\mathrm{AdS}_5\times \mathrm{S}^5$ supercoset sigma model is given by
\begin{align}
\text{ ii)}\qquad   (A|_{z=p},\partial_{z} A|_{z=p})&\in \mathfrak{su}(2,2|4)_{p,R_{n_p}}\qquad (p\in \mathfrak{p})
\,.
\label{eq:A-bsol-super-hYB}
\end{align}
The subscript $n_p$ of $R$ denotes the label of the poles as $\{n_1\,, n_i\,, n_{-1}\,, n_{-i}\}\equiv \{1,2,3,4\}$\,, 
and $\mathfrak{su}(2,2|4)_{p, R_{n_p}}$ is defined as
\begin{align}
   \mathfrak{su}(2,2|4)_{p, R_{n_p}}\equiv
   \left\{\begin{array}{ll}
   \left\{\left(p\, \eta\,R_{n_p}(x),x\right)\,|\,x\in\mathfrak{su}(2,2|4)\right\} \qquad &\mbox{for }p=\pm1\\
   \left\{\left(p\, \eta\,R_{n_p}(x),x\right)\,|\,x\in\mathfrak{su}(2,2|4)^{\mathbb{C}}\right\} & \mbox{for }p=\pm i\,.
   \end{array}\right.
    \label{hYB-DD-super}
\end{align}
Here the linear operators $R_k:\mathfrak{g}^{\mathbb{C}}\to \mathfrak{g}^{\mathbb{C}}\,(k=1,2,3,4)$ are 
\begin{align}
     R_{k}\equiv f_s^{k-1}\circ R \circ f_s^{-(k-1)}\,,
     \label{def:Rk}
\end{align}
where the linear $R$-operator $R:\mathfrak{su}(2,2|4)^{\mathbb{C}}\to \mathfrak{su}(2,2|4)^{\mathbb{C}}$ 
is a solution to the hCYBE for $\mathfrak{su}(2,2|4)^{\mathbb{C}}$\,, and $f_s:\mathfrak{su}(2,2|4)^{\mathbb{C}} 
\to \mathfrak{su}(2,2|4)^{\mathbb{C}}$ is a $\mathbb{Z}_4$-grading automorphism of $\mathfrak{g}^{\mathbb{C}}$\,. 
An explicit representation 
is given in (\ref{fs-def}). 
For this representation, one can show that the $R$-operator $R_k$ also satisfies the hCYBE (\ref{eq:hCYBE}) for $\mathfrak{su}(2,2|4)^{\mathbb{C}}$ if $R$ is a solution to the equation (\ref{eq:hCYBE}). 
Therefore, the boundary conditions (\ref{eq:A-bsol-super-hYB}) can be taken as solutions to the boundary equations of motion (\ref{eq:beom-sol}).

\subsubsection*{Lax form}

Similarly to the symmetric coset sigma model case, 
let us take $\hat{g}$ at each pole of the twist function (\ref{eq:twist-super}) as
\begin{align}
\begin{split}
    \hat{g}(\tau,\sigma,z)|_{z=1}&=g_1(\tau,\sigma)\,,\qquad \hat{g}(\tau,\sigma,z)|_{z=i}=g_2(\tau,\sigma)\,,\\
    \hat{g}(\tau,\sigma,z)|_{z=-1}&=g_3(\tau,\sigma)\,,\qquad \hat{g}(\tau,\sigma,z)|_{z=-i}=g_4(\tau,\sigma)
    \,,
    \end{split}
\end{align}
where $g_k \in SU(2,2|4)\,\,(k=1,3)$\,, $g_k \in SU(2,2|4)^{\mathbb{C}}\,\,(k=2,4)$\,,
such that 
\begin{align}
\tilde{\tau}g_{2}=g_{4}\,.
\end{align}
The associated left-invariant currents are defined as
\begin{align}
    j_1& \equiv g_1^{-1}dg_1\,,\qquad j_2 \equiv g_2^{-1}dg_2\,,\qquad
    j_3 \equiv g_3^{-1}dg_3\,,\qquad j_4 \equiv g_4^{-1}dg_4
    \,,
\end{align}
and the relations between the gauge field $A$ and the Lax pair $\cL$ at each pole are written as
\begin{align}
\begin{split}
    A|_{z=1}=-dg_1 g^{-1}_1&+{\rm Ad}_{g_1}\cL|_{z=1}\,,\qquad
    A|_{z=i}=-dg_2g^{-1}_2+{\rm Ad}_{g_2}\cL|_{z=i}\,,\\
    A|_{z=-1}=-dg_3 g^{-1}_3&+{\rm Ad}_{g_3}\cL|_{z=-1}\,,\qquad
    A|_{z=-i}=-dg_4g_4^{-1}+{\rm Ad}_{g_4}\cL|_{z=-i}\,.
    \label{eq:A-Lax-super}
    \end{split}
\end{align}
From the zero structure of the twist function (\ref{eq:twist-super}),
we suppose the following ansatz for the Lax pair as
\begin{align}
    \cL&=\left(z^{-1}\,V_{+}^{[-1]}+V_{+}^{[0]}+z\,V_{+}^{[1]}+z^2\,V_{+}^{[2]}\right)d\sigma^+\no\\
    &\quad +\left(z^{-2}\,V_{-}^{[-2]}+z^{-1}\,V_{-}^{[-1]}+V_{-}^{[0]}+z\,V_{-}^{[1]}\right)d\sigma^{-}\,,\label{eq:Lax-ansatz-super}
\end{align}
where $V_{\pm}^{[n]}\,(n=-1,0,1)\,,V_{\pm}^{[\pm2]}:\cM\to \mathfrak{su}(2,2|4)$ are smooth functions
such that $\tau\cL=\mu_{\rm t}^{*}\cL$.
As we will see later, the ansatz (\ref{eq:Lax-ansatz-super}) works well for both solutions to the boundary equations of motion.
Note that the above ansatz (\ref{eq:Lax-ansatz-super}) is not the only possible choice.
One may consider other ansatz corresponding to the pure spinor formalism by following \cite{CY}.

\subsection*{i) the $\mathrm{AdS}_5\times\mathrm{S}^5$ supercoset sigma model}

Let us reproduce the GS action of the $\mathrm{AdS}_5\times \mathrm{S}^5$ supercoset sigma model from the 4D CS action (\ref{4dcs}).

\medskip

The boundary conditions (\ref{eq:A-bsol-super}) lead to 
\begin{align}
\begin{split}
    j_{1,\pm}&=V_{\pm}^{[0]}+V_{\pm}^{[\pm2]}+V_{\pm}^{[1]}+V_{\pm}^{[-1]}\,,\\
    j_{2,\pm}&=V_{\pm}^{[0]}-V_{\pm}^{[\pm2]}+ i\,V_{\pm}^{[1]}- i\,V_{\pm}^{[-1]}\,,\\ 
    j_{3,\pm}&=V_{\pm}^{[0]}+V_{\pm}^{[\pm2]}-V_{\pm}^{[1]}-V_{\pm}^{[-1]}
    \,,\\ 
    j_{4,\pm}&=V_{\pm}^{[0]}-V_{\pm}^{[\pm2]}
    - i\,V_{\pm}^{[1]}+i\,V_{\pm}^{[-1]}\,.\label{j=v}
    \end{split}
\end{align}
By solving these equations with respect to $V_{\pm}^{[n]}$\,, we obtain 
\begin{align}
\begin{split}
    V_{\pm}^{[0]}&=\frac{ j_{1,\pm}+j_{2,\pm}+j_{3,\pm}+j_{4,\pm} }{4}\,,\qquad 
    V_{\pm}^{[\pm2]}=\frac{ j_{1,\pm}-j_{2,\pm}+j_{3,\pm}-j_{4,\pm} }{4}\,,\\
    V_{\pm}^{[1]}&=\frac{ j_{1,\pm}-i\,j_{2,\pm}-j_{3,\pm}+i\,j_{4,\pm} }{4}\,,\qquad 
    V_{\pm}^{[-1]}=\frac{ j_{1,\pm}+i\,j_{2,\pm}-j_{3,\pm}-i\,j_{4,\pm} }{4}\,.
    \label{eq:V-ex}
    \end{split}
\end{align}
Then, $\text{res}_{p}(\varphi_{\rm str}\,\mathcal{L})\,(p\in \mathfrak{p})$ are evaluated as
\begin{align}
\begin{split}
    \text{res}_{\pm 1}(\varphi_{\rm str}\,\mathcal{L})&=\frac{1}{8}\Bigl(j_{1,+}-(1\pm i)j_{2,+}+j_{3,+} -(1\mp i)j_{4,+}\Bigr)d\sigma^{+}\\
    &\qquad+\frac{1}{8}\Bigl(-j_{1,-}+(1\mp i) j_{2,-}-j_{3,-}+(1\pm i)j_{4,-}\Bigr)d\sigma^{-}\,,\\
    \text{res}_{\pm i}(\varphi_{\rm str}\, \mathcal{L})&=\frac{1}{8}\Bigl(-(1\mp i)j_{1,+}+ j_{2,+}-(1\pm i)j_{3,+}+j_{4,+}\Bigr)d\sigma^{+}\\
    &\qquad +\frac{1}{8}\Bigl((1\pm i)j_{1,-}-  j_{2,-}+(1\mp i)j_{3,-}-j_{4,-}\Bigr)d\sigma^{-}
    \,.
    \end{split}
    \label{eq:res_super}
\end{align}
Note that the set $\{\text{res}_{\pm 1}(\varphi_{\rm str}\,\mathcal{L}), \text{res}_{\pm i}(\varphi_{\rm str}\,\mathcal{L})  \}$ 
is invariant under a cyclic permutation of $g_{k}\,(k=1,\dots,4)$.
This fact indicates that the associated $2$D action also has the same symmetry.
In fact, by using (\ref{eq:res_super}), we obtain the $2$D action
\begin{align}
    S[g_{k}]
    &=\frac{1}{16}\int_{\cM}\Str\biggl[\,\sum_{\sigma\in S^{4}}\Bigl(j_{\sigma(1),+}-(1+ i)j_{\sigma(2),+}+j_{\sigma(3),+} -(1- i)j_{\sigma(4),+}\Bigr)j_{\sigma(1),-}\no\\
    &\qquad-\Bigl(-j_{\sigma(1),-}+(1- i) j_{\sigma(2),-}-j_{\sigma(3),-}+(1+ i)j_{\sigma(4),-}\Bigr)j_{\sigma(1),+}\biggr]d\sigma^+\wedge d\sigma^-\,,
    \label{eq:AdS5action-pre}
\end{align}
where $\sigma \in S^{4}$ is a cyclic permutation of the set $\{1,2,3,4\}$\,.
The action (\ref{eq:AdS5action-pre}) is clearly invariant under the cyclic permutations of $j_{k}$\,.

\medskip

Furthermore, we impose relations among $j_{k}\,(k=1\,,\dots\,,4)$\,.
From the cyclic symmetry of the 2D action (\ref{eq:AdS5action-pre}),
we can require the relation 
\begin{align}
  j_k=f_s^{k-1}(j)\qquad   (k=1\,,\dots\,,4)\,,
  \label{jk=fj}
\end{align}
where $j\in \mathfrak{su}(2,2|4)$ is the left-invariant current for $g \in SU(2,2|4)$\,, and the map $f_s:\mathfrak{su}(2,2|4)^{\mathbb{C}}\to \mathfrak{su}(2,2|4)^{\mathbb{C}}$ is an automorphism of $\mathfrak{su}(2,2|4)$ satisfying the $\mathbb{Z}_4$-grading property $f_s^4=\text{Id}$\,. 
As is well known, the superalgebra $\mathfrak{su}(2,2|4)$ has the following decomposition into vector subspaces with respect to the $\mathbb{Z}_4$-grading structure:
\begin{align}
    \mathfrak{g}=\mathfrak{g}^{(0)}\oplus \mathfrak{g}^{(1)}\oplus \mathfrak{g}^{(2)}\oplus \mathfrak{g}^{(3)}\,,
\end{align}
where $\mathfrak{g}^{(0)}\oplus \mathfrak{g}^{(2)}$ and $\mathfrak{g}^{(1)}\oplus \mathfrak{g}^{(3)}$ are the bosonic and  fermionic parts of $\mathfrak{su}(2,2|4)$\,, respectively, and $\mathfrak{g}^{(0)}$ is identified with a bosonic subgroup $\mathfrak{so}(1,4)\times \mathfrak{so}(5)$ .
The commutation relations of $\mathfrak{g}^{(m)}$ satisfy
\begin{align}
    [\mathfrak{g}^{(m)},\mathfrak{g}^{(n)}]\subset \mathfrak{g}^{(k)}\qquad (m+n=k\quad \text{mod}\,\;4)\,.
\end{align}
In order to obtain the GS action, let us take the $\mathbb{Z}_4$-grading automorphism $f_s$ such that each subspace $\mathfrak{g}^{(k)}\,(k=0,1,2,3)$ is the eigenspace of $f_s$ satisfying
\begin{align}
    f_s(\mathfrak{g}^{(k)})=i^{k}\mathfrak{g}^{(k)}\,.\label{fs-def}
\end{align}
Note that after taking a supermatrix realization of $\mathfrak{su}(2,2|4)$\,, we can write down the explicit expression of $f_s$ (For the details, see \cite{Arutyunov:2009ga}).

\medskip

The additional condition (\ref{jk=fj}) enables us to express the functions $V^{[n]}$ in terms of 
the $\mathbb{Z}_4$-graded components $j^{(k)}_{\pm}\in \mathfrak{g}^{(k)}$\,.
In fact, by using (\ref{jk=fj}) and (\ref{fs-def}), the left-invariant currents in (\ref{jk=fj}) are rewritten as
\begin{align}
\begin{split}
    j_{1,\pm}&=j_{\pm}^{(0)}+j_{\pm}^{(1)}+j_{\pm}^{(2)}+j_{\pm}^{(3)}\,,\qquad 
    j_{2,\pm}=j_{\pm}^{(0)}+i\,j_{\pm}^{(1)}-j_{\pm}^{(2)}-i\,j_{\pm}^{(3)}\,,\\
    j_{3,\pm}&=j_{\pm}^{(0)}-j_{\pm}^{(1)}+j_{\pm}^{(2)}-j_{\pm}^{(3)}\,,\qquad 
    j_{4,\pm}=j_{\pm}^{(0)}-i\,j_{\pm}^{(1)}-j_{\pm}^{(2)}+i\,j_{\pm}^{(3)}\,.\label{eq:Z4-conf}
    \end{split}
\end{align}
Then, by substituting (\ref{eq:Z4-conf}) into (\ref{eq:V-ex}), the functions $V^{[n]}$ are given by
\begin{align}
    V_{\pm}^{[0]}&=j_{\pm}^{(0)}\,,\qquad 
    V_{\pm}^{[\pm2]}=j_{\pm}^{(2)}\,,\qquad 
    V_{\pm}^{[1]}=j_{\pm}^{(1)}\,,\qquad 
    V_{\pm}^{[-1]}=j_{\pm}^{(3)}\,.
    \label{V-sol-super}
\end{align}
From this result, we immediately obtain the Lax pair 
\begin{align}
    \cL&=\left(z^{-1}\,j_{+}^{(3)}+j_{+}^{(0)}+z\,j_{+}^{(1)}+z^2\,j_{+}^{(2)}\right)d\sigma^+\no\\
    &\quad +\left(z^{-2}\,j_{-}^{(2)}+z^{-1}\,j_{-}^{(3)}+j_{-}^{(0)}+z\,j_{-}^{(1)}\right)d\sigma^{-}\,.
    \label{eq-BPR-Lax}
\end{align}
The expression (\ref{eq-BPR-Lax}) is precisely the same as the Lax pair constructed in \cite{Bena:2003wd}.

\medskip

Next, let us evaluate the 2D action (\ref{eq:AdS5action-pre}).
By using (\ref{V-sol-super}), we can see that the contribution to the 2D action from each pole is 
identical, namely, 
\begin{align}
    \Str\left( \text{res}_{p}(\varphi_{\rm str}\mathcal{L})\wedge g^{-1}_{p}d g_{p}\right)
    =\frac{1}{2} \Str\left( j_{-}d_{+}(j_{+})\right) d\sigma^+\wedge d\sigma^-\,,
\end{align}
where $d_{\pm}$ are the linear combinations of the projection operators $P_{(i)}$ like 
\begin{align}
    d_{\pm}=\pm P_{(1)}+2P_{(2)}\mp P_{(3)}\,. 
\end{align}
This fact comes from the cyclic symmetry of the 2D action (\ref{eq:AdS5action-pre}).
As a result, we obtain 
\begin{align}
    S[g_{i}]
    &=\int_{\cM}\Str\left( j_{-}d_{+}(j_{+})\right) d\sigma^+\wedge d\sigma^-\,.
\end{align}
This is nothing but the Metsaev-Tseytlin action of the $\mathrm{AdS}_5\times \mathrm{S}^5$ 
supercoset sigma model \cite{Metsaev:1998it}.

\subsection*{ii) homogeneous YB deformations}

Let us next discuss homogeneous YB deformations of the $\mathrm{AdS}_5\times \mathrm{S}^5$ supercoset sigma model \cite{KMY1}.

\medskip

We consider the boundary condition (\ref{eq:A-bsol-super-hYB}).
To avoid confusion of notations, we will replace the functions $V_{\pm}^{[n]}$ appeared in the Lax pair (\ref{eq:Lax-ansatz-super}) with  $\overline{V}_{\pm}^{[n]}(\in\mathfrak{su}(2,2|4))$.
Then, from the boundary condition (\ref{eq:A-bsol-super-hYB}), we obtain the relations
\begin{align}
\begin{split}
    j_{1,\pm}&=\overline{V}_{\pm}^{[0]}+(1\mp 2\eta R_{g_1})\overline{V}_{\pm}^{[\pm2]}
    +(1- \eta R_{g_1})\overline{V}_{\pm}^{[1]}+(1+ \eta R_{g_1})\overline{V}_{\pm}^{[-1]}\,,\\
    j_{2,\pm}&=\overline{V}_{\pm}^{[0]}-(1\mp 2 \eta R_{g_2})\overline{V}_{\pm}^{[\pm2]}
     +i\,(1- \eta R_{g_2})V_{\pm}^{[1]}-i\,(1+ \eta R_{g_2})\overline{V}_{\pm}^{[-1]}\,,\\ 
    j_{3,\pm}&=\overline{V}_{\pm}^{[0]}+(1\mp 2\eta R_{g_3})\overline{V}_{\pm}^{[\pm2]}
    -(1- \eta R_{g_3})\overline{V}_{\pm}^{[1]}-(1+ \eta R_{g_3})\overline{V}_{\pm}^{[-1]}
    \,,\\ 
    j_{4,\pm}&=\overline{V}_{\pm}^{[0]}-(1\mp 2 \eta R_{g_4})\overline{V}_{\pm}^{[\pm2]}
     -i\,(1- \eta R_{g_4})\overline{V}_{\pm}^{[1]}+i\,(1+ \eta R_{g_4})\overline{V}_{\pm}^{[-1]}\,,  
\label{jbVeq}
    \end{split}
\end{align}
where the dressed $R$-operator $R_{g_{k}}\,(k=1\,,\dots \,, 4)$ is defined as
\begin{align}
    R_{g_{k}}\equiv\Ad_{g_k}^{-1}\circ R_k \circ \Ad_{g_k}\,.
\end{align}
By introducing the linear operator
\begin{align}
   \gR_{\bgg}^{(p)}= \frac{1}{4}\left(R_{g_1}+i^{p}R_{g_2}+i^{2p}R_{g_3}+i^{3p}R_{g_4}\right)\,,\label{Rg-sum}
\end{align}
the equations (\ref{jbVeq}) are rewritten as 
\begin{align}
\begin{split}
  V^{[0]}_{\pm}&=
    \overline{V}_{\pm}^{[0]}\mp 2\eta\,\gR_{\bgg}^{(2)}\,\overline{V}_{\pm}^{[\pm 2]}-\eta\,\gR_{\bgg}^{(1)}\,\overline{V}_{\pm}^{[1]}  +\eta\,\gR_{\bgg}^{(3)}\,\overline{V}_{\pm}^{[-1]}\,,\\%%%%%%%%%%%%%%%%%%%%%%%%%%%%%%%%%
   V^{[1]}_{\pm}&=
  \mp2\eta \,\gR_{\bgg}^{(1)}\,\overline{V}_{\pm}^{[\pm 2]}+\left(1-\eta\,\gR_{\bgg}^{(0)}\right)\overline{V}_{\pm}^{[1]}
     +\eta\,\gR_{\bgg}^{(2)}\,\overline{V}_{\pm}^{[-1]}\,,\\%%%%%%%%%%%%%%%%%%%%%%%%%%%%%%%%%%%
   V^{[\pm2]}_{\pm}&=
    \left(1\mp2\eta \,\gR_{\bgg}^{(0)}\right)\overline{V}_{\pm}^{[\pm 2]}-\eta\,\gR_{\bgg}^{(3)}\,\overline{V}_{\pm}^{[1]} +\eta\,\gR_{\bgg}^{(1)}\,\overline{V}_{\pm}^{[-1]}\,,\\%%%%%%%%%%%%%%%%%%%%%%%%%%%%%%%%%%
   V^{[-1]}_{\pm}&=
    \mp2\eta\,\gR_{\bgg}^{(3)}\,\overline{V}_{\pm}^{[\pm 2]} -\eta\,\gR_{\bgg}^{(2)}\,\overline{V}_{\pm}^{[1]}  +\left(1+\eta\,\gR_{\bgg}^{(0)}\right)\overline{V}_{\pm}^{[-1]}\,,
        \label{eq:j=V-2}
    \end{split}
\end{align}
where the functions $V^{[n]}_{\pm}$ take the expressions (\ref{eq:V-ex}).
Since the operator $\gR_{\bgg}^{(p)}$ is skew-symmetric, the equations (\ref{eq:j=V-2}) for $\overline{V}^{[n]}_{\pm}$ can be uniquely solved and the associated 2D action can also be written down.
However, the resulting 2D action has a rather complex form, and so instead of giving its explicit expression, we will only show that the associated 2D action is invariant under the cyclic permutation of $g_k\,(k=1,\dots,4)$\,.

\medskip

For this purpose, let us define the map
\begin{align}
    \sfP:g_k\mapsto g_{k+1}\,,
\end{align}
Under this transformation, the linear operator $\gR_{\bgg}^{(p)}$ in (\ref{Rg-sum}) 
and the functions $V^{[n]}_{\pm}$ in (\ref{eq:V-ex}) are transformed as
\begin{align}
    \sfP\left(  \gR_{\bgg}^{(p)} \right)=i^{3p}\,\gR_{\bgg}^{(p)} \,,\label{PRgp}
\end{align}
and
\begin{align}
\begin{split}
   \sfP\left( V^{[0]}_{\pm} \right)&= V^{[0]}_{\pm}\,,\qquad  
      \sfP\left( V^{[\pm 2]}_{\pm} \right)= -V^{[\pm 2]}_{\pm}\,,\\
      \sfP\left( V^{[1]}_{\pm} \right)&= i\,V^{[1]}_{\pm}\,,\qquad 
       \sfP\left( V^{[-1]}_{\pm} \right)= -i\,V^{[-1]}_{\pm}\,.\label{PV=V}
       \end{split}
\end{align}
From the transformation rules in (\ref{PRgp}) and (\ref{PV=V}), and the equations (\ref{eq:j=V-2}), the functions $\overline{V}^{[n]}_{\pm}$ follow the same transformation rules as the functions $V^{[n]}_{\pm}$\,,
\begin{align}
\begin{split}
   \sfP\left( \overline{V}^{[0]}_{\pm} \right)&= \overline{V}^{[0]}_{\pm}\,,\qquad 
      \sfP\left( \overline{V}^{[\pm 2]}_{\pm} \right)= -\overline{V}^{[\pm 2]}_{\pm}\,,\\
      \sfP\left( \overline{V}^{[1]}_{\pm} \right)&= i\,\overline{V}^{[1]}_{\pm}\,,\qquad  \sfP\left( \overline{V}^{[-1]}_{\pm} \right)= -i\,\overline{V}^{[-1]}_{\pm}\,.
      \end{split}
\end{align}
This fact indicates that the residues $\text{res}_{p}(\varphi_{\rm str}\,\mathcal{L})\,(p\in \mathfrak{p})$ satisfy
\begin{align}
    \sfP\left( \text{res}_{p}(\varphi_{\rm str}\,\mathcal{L})\right)= \text{res}_{p+1}(\varphi_{\rm str}\,\mathcal{L})\,,
\end{align}
where $\text{res}_{p}(\varphi_{\rm str}\,\mathcal{L})\,(p\in \mathfrak{p})$ is given by
\begin{align}
    \text{res}_{p}(\varphi_{\rm str}\,\mathcal{L})&=\frac{1}{4}\Bigl(i^{n_p-1}\, \overline{V}_{+}^{[1]}+2i^{2(n_p-1)}\overline{V}_{+}^{[2]}-i^{3(n_p-1)} \overline{V}_{+}^{[-1]} \Bigr)d\sigma^{+}\\
    &\qquad+\frac{1}{4}\Bigl(i^{n_p-1}\, \overline{V}_{-}^{[1]}-2i^{2(n_p-1)}\overline{V}_{-}^{[2]}-i^{3(n_p-1)} \overline{V}_{-}^{[-1]}\Bigr)d\sigma^{-}
    \,.
\end{align}
Therefore, the associated 2D action (\ref{2d action}) is invariant under the permutation of $g_k\,(k=1,\dots,4)$\,.

\medskip

Thanks to the cyclic symmetry of the 2D action, we can require an additional condition
\begin{align}
  g_k=F_s^{k-1}(g)\qquad   (k=1\,,\dots\,,4)\,,
\end{align}
where $g\in SU(2,2|4)$\,, and the map $F_{s}: SU(2,2|4)\to SU(2,2|4)$ is an automorphism of $SU(2,2|4)$ satisfying $F_{s}^4=1$\,.
As in the symmetric coset case, let us take $F_s$ so as to be induced by $f_s$ defined in (\ref{fs-def}).
More concretely, when a parameterization of an element $g\in SU(2,2|4)$ is taken as
\begin{align}
\begin{split}
    g=\exp(\sum_{k=0}^{3}X^{A_k} T_{A_k}^{(k)})\,,\qquad
    T^{(k)}_{A_k}\in \mathfrak{g}^{(k)}\qquad (A_k=1,\dots,\text{dim}\,\mathfrak{g}^{(k)})\,,
\end{split}
\end{align}
the automorphism $F_s$ is defined as
\begin{align}
    F_s(g)&\equiv\exp(\sum_{k=0}^{3}X^{A_k} f_s( T_{A_k}^{(k)}))=\exp(\sum_{k=0}^{3}i^{k}\,X^{A_k} T_{A_k}^{(k)})\,.
\end{align}
Here, $X^{A_k}$ are functions of $\tau$ and $\sigma$\,.
By definition, $F_s$ is an automorphism of $SU(2,2|4)$ with the $\mathbb{Z}_4$-grading property.

\medskip

Then, as shown in appendix \ref{sec:proofRgsuper}, the dressed $R$-operator $R_{g_k}$ 
that act on the generators of $\mathfrak{su}(2,2|4)$ should satisfy
\begin{align}
   P^{(m)}\circ R_{g_k}\circ P^{(n)}=i^{(m-n)(k-1)}P^{(m)}\circ R_{g}\circ P^{(n)}\,.\label{Rgk-Rg}
\end{align}
This relation indicates 
\begin{align}
    P^{(m)}\circ   \gR_{\bgg}^{(p)} \circ P^{(n)}=
    \begin{cases}
     P^{(m)}\circ R_{g}\circ P^{(n)}\qquad &m-n+p=0\,({\rm mod}\,4)\\
    0\qquad &m-n+p\neq 0\,({\rm mod}\,4)
    \end{cases}
    \,.
    \label{eq:PRP}
\end{align}
By using the relation (\ref{eq:PRP}), the equations in (\ref{eq:j=V-2}) can be solved as 
\begin{align}
    \overline{V}_{\pm}^{[0]}=J_{\pm}^{(0)}\,,\qquad \overline{V}_{\pm}^{[\pm2]}=J_{\pm}^{(2)}\,,\qquad 
    \overline{V}_{\pm}^{[1]}=J_{\pm}^{(1)}\,,\qquad \overline{V}_{\pm}^{[-1]}=J_{\pm}^{(3)}\,,
        \label{V-sol-super-hYB}
\end{align}
where the deformed current $J_{\pm}$ is defined as 
\begin{align}
   J_{\pm}\equiv \frac{1}{1\mp\eta R_{g}\circ d_{\pm}}j_{\pm}\,.
\end{align}
Thus the Lax pair is given by  
\begin{align}
    \cL&=\left(z^{-1}\,J_{+}^{(3)}+J_{+}^{(0)}+z\,J_{+}^{(1)}+z^2\,J_{+}^{(2)}\right)d\sigma^+\no\\
    &\quad +\left(z^{-2}\,J_{-}^{(2)}+z^{-1}\,J_{-}^{(3)}+J_{-}^{(0)}+z\,J_{-}^{(1)}\right)d\sigma^{-}\,.
    \label{eq-hYB-sLax}
\end{align}
This is nothing but the Lax pair of homogeneous YB deformations of 
the $\mathrm{AdS}_5\times \mathrm{S}^5$ supercoset sigma model \cite{KMY1}.

\medskip

Next, let us derive the associated 2D action.
By using (\ref{V-sol-super-hYB}), we find that the contribution to the 2D action from each pole is identical, 
namely, 
\begin{align}
    \Str\left( \text{res}_{p}(\varphi_{\rm str}\,\mathcal{L})\wedge g^{-1}_{p}d g_{p}\right)
    =\frac{1}{2} \Str\left( j_{-}d_{+}(J_{+})\right) d\sigma^+\wedge d\sigma^-\,.
\end{align}
As a result, we obtain 
\begin{align}
    S[g]
    &=\int_{\cM}\Str\left( j_{-}d_{+}(J_{+})\right) d\sigma^+\wedge d\sigma^-\,.
    \label{eq:hYB-super-action}
\end{align}
This action (\ref{eq:hYB-super-action}) is precisely the same as that of homogeneous YB deformations 
of the $\mathrm{AdS}_5\times \mathrm{S}^5$ supercoset sigma model \cite{KMY1}.

\section{Conclusion and Discussion}

In this paper, we have generalized the preceding result on the PCM to the case of the symmetric coset sigma model. 
By employing the same twist function in the rational description, we have specified boundary conditions 
which lead to the symmetric coset sigma model and the homogeneous YB-deformed relatives. 
The same analysis is applicable for the AdS$_5\times$S$^5$ supercoset sigma model. As a result, homogeneous YB-deformations 
of the AdS$_5\times$S$^5$ supercoset sigma model have been derived from the 4D CS theory as boundary conditions. 
In order to discuss the AdS$_5\times$S$^5$ superstring beyond the sigma model, we have to take the Virasoro conditions into account by following the seminal work \cite{CS} in the present formulation. This is one of the most significant issues and 
the result will be reported in another place \cite{Fukushima2}. 

\medskip

There are some open questions. It is well known that homogeneous YB deformations with abelian classical $r$-matrices 
can be seen as twisted boundary conditions \cite{Frolov,AAF,MY-LM,MY-MR,Osten:2016dvf} via non-local gauge transformations. 
It is interesting to consider the interpretation of this fact from the viewpoint of the 4D CS theory. 
It is also significant to understand how to realize the sine-Gordon model from the 4D CS theory. 
The sine-Gordon model can be reproduced from the $O(3)$ NLSM via the Pohlmeyer reduction at the classical level. 
Hence it would be nice to study how the Pohlmeyer reduction works in the context of the 4D CS theory. 

\medskip

It is also interesting to study the $\eta$-deformation based on the modified classical YB equation as well, 
though we have discussed only the homogeneous YB-deformations. 
We will report the result in another place \cite{Fukushima2}.

\subsection*{Acknowledgments}

The work of J.S.\ was supported in part by Ministry of Science and Technology (project no. 108-2811-M-002-528), 
National Taiwan University.
The works of K.Y.\ was supported by the Supporting Program for Interaction-based Initiative 
Team Studies (SPIRITS) from Kyoto University, and JSPS Grant-in-Aid for Scientific Research (B) 
No.\,18H01214. This work is also supported in part by the JSPS Japan-Russia Research 
Cooperative Program.

\appendix

\section*{Appendix}

\section{Relations for dressed $R$-operators}

Here we shall prove the relations (\ref{eq:RgRtg}) and (\ref{Rgk-Rg}) that dressed $R$-operators should satisfy.

\subsection{$\mathbb{Z}_2$-grading case}\label{sec:proofRgRtg}

Let us first give a proof of the relation (\ref{eq:RgRtg}) for a dressed $R$-operator.

\medskip

To begin with, we examine how a dressed $R$-operator $R_{g}$ acts on the generators.
The adjoint operation with a group element $g$ on the generators $\gP_{\cha}$ and $\gJ_{\ha}$ is expressed as 
\begin{align}
    \Ad_{g}(\gP_{\cha})&=[\Ad_g]_{\cha}{}^{\chb}\,\gP_{\chb}+[\Ad_g]_{\cha}{}^{\ha}\,\gJ_{\ha}\,,\qquad
     \Ad_{g}(\gJ_{\ha})=[\Ad_g]_{\ha}{}^{\cha}\,\gP_{\cha}+[\Ad_g]_{\ha}{}^{\hb}\,\gJ_{\hb}\,.
\end{align}
Then the action of $R_g$ on $\gP_{\cha}$ is evaluated as
\begin{align}
    R_g(\gP_{\cha})&=\Ad_{g^{-1}}\circ R([\Ad_g]_{\cha}{}^{\chb}\,\gP_{\chb}+[\Ad_g]_{\cha}{}^{\ha}\,\gJ_{\ha} )\no\\
    &=\Ad_{g^{-1}} \Bigl([\Ad_g]_{\cha}{}^{\chb}\,R_{\chb}{}^{\chc}\,\gP_{\chc}+[\Ad_g]_{\cha}{}^{\chb}\,R_{\chb}{}^{\ha}\,\gJ_{\ha}
    +[\Ad_g]_{\cha}{}^{\ha}R_{\ha}{}^{\chb}\,\gP_{\chb}+[\Ad_g]_{\ha}{}^{\cha}R_{\cha}{}^{\chb}\,\gJ_{\chb} \Bigr)\no\\
    &= [\Ad_g]_{\cha}{}^{\chb}\,R_{\chb}{}^{\chc}[\Ad_{g^{-1}}]_{\chc}{}^{\chd}\,\gP_{\chd}+ [\Ad_g]_{\cha}{}^{\chb}\,R_{\chb}{}^{\chc}[\Ad_{g^{-1}}]_{\chc}{}^{\ha}\,\gJ_{\ha}\no\\
    &\quad+[\Ad_g]_{\cha}{}^{\chb}\,R_{\chb}{}^{\ha}[\Ad_{g^{-1}}]_{\ha}{}^{\chc}\,\gP_{\chc}+[\Ad_g]_{\cha}{}^{\chb}\,R_{\chb}{}^{\ha}[\Ad_{g^{-1}}]_{\ha}{}^{\hb}\,\gJ_{\hb}\no\\
    &\quad+[\Ad_g]_{\cha}{}^{\ha}R_{\ha}{}^{\chb}[\Ad_{g^{-1}}]_{\chb}{}^{\chc}\,\gP_{\chc}+[\Ad_g]_{\cha}{}^{\ha}R_{\ha}{}^{\chb}[\Ad_{g^{-1}}]_{\chb}{}^{\hb}\,\gJ_{\hb}\no\\
    &\quad+[\Ad_g]_{\cha}{}^{\ha}R_{\ha}{}^{\hb}[\Ad_{g^{-1}}]_{\hb}{}^{\chb}\,\gP_{\chb} +[\Ad_g]_{\cha}{}^{\ha}R_{\ha}{}^{\hb}[\Ad_{g^{-1}}]_{\hb}{}^{\hc}\,\gJ_{\hc} \,.
    \label{eq:Rg-ex}
\end{align}

\medskip

Next, let us see the adjoint actions of $\tilde{g}$\,, which is related to $g$ through the $\mathbb{Z}_2$-grading 
automorphism (\ref{Aut-ad}).
By using the Campbell-Baker-Hausdorff formula and the $\mathbb{Z}_2$-grading property of $\mathfrak{g}$, we can obtain
\begin{align}
    \Ad_{\tilde{g}}\left(\gP_{\cha}\right)&=\sum_{n=0}^{\infty}\frac{1}{n!}(-{\rm ad}_{X^{\chb}\gP_{\chb} }+{\rm ad}_{X^{\hb}\gJ_{\hb}})^{n}\left(\gP_{\cha}\right)\no\\
    &=\sum_{n=0}^{\infty}\frac{1}{n!}\biggl((\text{even number of }{\rm ad}_{X^{\chb}\gP_{\chb} })-(\text{odd number of }{\rm ad}_{X^{\chb}\gP_{\chb} })\biggr)\left(\gP_{\cha}\right)\no\\
     &=[\Ad_{g}]_{\cha}{}^{\chb}\,\gP_{\chb}-[ \Ad_{g}]_{\cha}{}^{\hb}\,\gJ_{\hb}\,,\\
    \Ad_{\tilde{g}}\left(\gJ_{\ha}\right)&=\sum_{n=0}^{\infty}\frac{1}{n!}(-{\rm ad}_{X^{\chb}\gP_{\chb} }+{\rm ad}_{X^{\hb}\gJ_{\hb}})^{n}\left(\gJ_{\ha}\right)\no\\
    &=\sum_{n=0}^{\infty}\frac{1}{n!}\biggl(-(\text{odd number of }{\rm ad}_{X^{\chb}\gP_{\chb} })+(\text{even number of }{\rm ad}_{X^{\chb}\gP_{\chb} })\biggr)\left(\gJ_{\ha}\right)\no\\
     &=-[\Ad_{g}]_{\ha}{}^{\chb}\,\gP_{\chb}+[ \Ad_{g}]_{\ha}{}^{\hb}\,\gJ_{\hb}\,.
\end{align} 
These results indicate that the adjoint action with $\tilde{g}$ is given by
\begin{align}
    \Ad_{\tilde{g}}(\gP_{\cha})&=[\Ad_{g}]_{\cha}{}^{\chb}\,\gP_{\chb}-[ \Ad_{g}]_{\cha}{}^{\ha}\,\gJ_{\ha}\,,\qquad
     \Ad_{\tilde{g}}(\gJ_{\ha})=-[\Ad_{g}]_{\ha}{}^{\cha}\,\gP_{\cha}+[ \Ad_{g}]_{\ha}{}^{\hb}\,\gJ_{\hb}\,.
\end{align}
Then, the action of $\tilde{R}_{\tilde{g}}$ on $\gP_{\cha}$ defined in (\ref{eq:tR}) is given by
\begin{align}
    \tilde{R}_{\tilde{g}}(\gP_{\cha})&=\Ad_{\tilde{g}^{-1}}\circ \tilde{R}([\Ad_g]_{\cha}{}^{\chb}\,\gP_{\chb}-[\Ad_g]_{\cha}{}^{\ha}\,\gJ_{\ha} )\no\\
    &=\Ad_{\tilde{g}^{-1}}\circ f^{-1} \circ R(-[\Ad_g]_{\cha}{}^{\chb}\,\gP_{\chb}-[\Ad_g]_{\cha}{}^{\ha}\,\gJ_{\ha} )\no\\
    &=\Ad_{\tilde{g}^{-1}} \Bigl([\Ad_g]_{\cha}{}^{\chb}\,R_{\chb}{}^{\chc}\,\gP_{\chc}-[\Ad_g]_{\cha}{}^{\chb}\,R_{\chb}{}^{\ha}\,\gJ_{\ha}
    +[\Ad_g]_{\cha}{}^{\ha}R_{\ha}{}^{\chb}\,\gP_{\chb}
    -[\Ad_g]_{\cha}{}^{\ha}R_{\ha}{}^{\chb}\,\gJ_{\chb} \Bigr)\no\\
    &= [\Ad_g]_{\cha}{}^{\chb}\,R_{\chb}{}^{\chc}[\Ad_{g^{-1}}]_{\chc}{}^{\chd}\,\gP_{\chd}- [\Ad_g]_{\cha}{}^{\chb}\,R_{\chb}{}^{\chc}[\Ad_{g^{-1}}]_{\chc}{}^{\ha}\,\gJ_{\ha}\no\\
    &\quad+[\Ad_g]_{\cha}{}^{\chb}\,R_{\chb}{}^{\ha}[\Ad_{g^{-1}}]_{\ha}{}^{\chc}\gP_{\chc}-[\Ad_g]_{\cha}{}^{\chb}\,R_{\chb}{}^{\ha}[\Ad_{g^{-1}}]_{\ha}{}^{\hb}\gJ_{\hb}\no\\
    &\quad+[\Ad_g]_{\cha}{}^{\ha}R_{\ha}{}^{\chb}[\Ad_{g^{-1}}]_{\chb}{}^{\chc}\,\gP_{\chc}-[\Ad_g]_{\cha}{}^{\ha}R_{\ha}{}^{\chb}[\Ad_{g^{-1}}]_{\chb}{}^{\hb}\,\gJ_{\hb}\no\\
    &\quad+[\Ad_g]_{\cha}{}^{\ha}R_{\ha}{}^{\hb}[\Ad_{g^{-1}}]_{\hb}{}^{\chb}\,\gP_{\chb} -[\Ad_g]_{\cha}{}^{\ha}R_{\ha}{}^{\hb}[\Ad_{g^{-1}}]_{\hb}{}^{\hc}\,\gJ_{\hc} \,.
        \label{eq:tRg-ex}
\end{align}
By using (\ref{eq:Rg-ex}) and (\ref{eq:tRg-ex}), we can obtain the relation (\ref{eq:RgRtg}).

\subsection{$SU(2,2|4)$ case}\label{sec:proofRgsuper}

Next, let us show that the action of the dressed $R$-operator $R_{g_k}\,(k=1,\dots,4)$ 
on the $\mathfrak{su}(2,2|4)$ generators satisfies the relation (\ref{Rgk-Rg}).

\medskip

As in the previous case, we can see that the adjoint action with $g_k$ on the generators of $\mathfrak{su}(2,2|4)$ 
is written as
\begin{align}
    \Ad_{g_k}\circ P^{(n)}&=\sum_{s=0}^{3}i^{(s-n)(k-1)}P^{(s)}\circ \Ad_g\circ P^{(n)}\,,\\
    P^{(m)}\circ\Ad_{g_k}^{-1}&=\sum_{r=0}^{3}i^{(m-r)(k-1)}P^{(m)}\circ \Ad_g^{-1}\circ P^{(r)}\,.
\end{align} 
By using these relations and the definition (\ref{def:Rk}) of $R_{g_{k}}$, the projected dressed $R$-operator 
$P^{(m)}\circ R_{g_k}\circ P^{(n)}$ can be expressed as
\begin{align}
   P^{(m)}\circ R_{g_k}\circ P^{(n)}&=  P^{(m)}\circ\Ad_{g_k}^{-1}\circ f_s^{k-1}\circ R\circ f_s^{-(k-1)}\circ \left(\sum_{s=0}^{3}i^{(s-n)(k-1)}P^{(s)}\circ \Ad_g\circ P^{(n)}\right)\no\\
   &=P^{(m)}\circ\Ad_{g_k}^{-1}\circ f_s^{k-1}\circ   \left(\sum_{s=0}^{3}i^{(s-n)(k-1)-s (k-1)}R\circ P^{(s)}\circ \Ad_g\circ P^{(n)}\right)\no\\
   &=P^{(m)}\circ\Ad_{g_k}^{-1}\circ   \left(\sum_{r,s=0}^{3}i^{-n(k-1)+r(k-1)}P^{(r)}\circ R\circ P^{(s)}\circ \Ad_g\circ P^{(n)}\right)\no\\
    &=\left(\sum_{r,s=0}^{3}i^{(m-n)(k-1)}P^{(m)}\circ\Ad_{g_k}^{-1}\circ P^{(r)}\circ R\circ P^{(s)}\circ \Ad_g\circ P^{(n)}\right)\no\\
    &=i^{(m-n)(k-1)}P^{(m)}\circ R_{g}\circ P^{(n)}\,.
\end{align}
Thus the relation (\ref{Rgk-Rg}) has been shown.

\section{Homogeneous bi-YB deformed sigma model}\label{sec:bi-YB}

In this appendix, let us derive the action of a homogeneous bi-YB deformed principal chiral model, 
which is a two-parameter generalization of homogeneous YB deformation.
In this case, we use the twist function (\ref{eq:twist-new}) which is the same as in the symmetric coset case.

\subsubsection*{Boundary condition}

A solution to the boundary equations of motion (\ref{eq:beom}) is given by 
\begin{align}
(A|_{z=1},\partial_{z} A|_{z=1})\in \mathfrak{g}_{R_{\sfR}}^{\mathbb{C}}\,,\qquad (A|_{z=-1},\partial_{z} A|_{z=-1})\in \mathfrak{g}_{R_{\sfL}}^{\mathbb{C}}\,,
\label{eq:A-bsol-bi}
\end{align}
where $\mathfrak{g}_{\sfR}^{\mathbb{C}}$ and $\mathfrak{g}_{\sfL}^{\mathbb{C}}$ are defined as
\begin{align}
    \mathfrak{g}_{R_{\sfR}}^{\mathbb{C}} \equiv \left\{\left(2\eta_{\sfR} R_{\sfR}(x),x\right)\,|\,x\in\mathfrak{g}^{\mathbb{C}}\right\}\,,\qquad    \mathfrak{g}_{R_{\sfL}}^{\mathbb{C}} \equiv 
\left\{\left(-2\eta_{\sfL} R_{\sfL}(x),x\right)\,|\,x\in\mathfrak{g}^{\mathbb{C}}\right\}\,.
    \label{def;gRL}
\end{align}
Here $\eta_{\sfR}$ and $\eta_{\sfL}$ are the deformation parameters, and $R_{\sfR}$ and $R_{\sfL}$ are 
linear $R$-operators satisfying the hCYBE (\ref{eq:hCYBE}).

\subsubsection*{Lax form}

Next, let us take $\hat{g}$ at each pole of the twist function (\ref{eq:twist-new}) as
\begin{align}
    \hat{g}(\tau,\sigma,z)|_{z=1}=g_{\sfR}(\tau,\sigma)\,,\qquad \hat{g}(\tau,\sigma,z)|_{z=-1}=g_{\sfL}(\tau,\sigma)\,,
\end{align}
where $g_{\sfR}\,, g_{\sfL} \in G^{\mathbb{C}}$ (rather than $G$)\,.
Then, the relation between the gauge field $A$ and the Lax pair $\cL$ at each pole is written as, respectively,     
\begin{align}
    A|_{z=1}=-dg_{\sfR}g^{-1}_{\sfR}+{\rm Ad}_{g_{\sfR}}\cL|_{z=1}\,,\qquad
    A|_{z=-1}=-dg_{\sfL}g^{-1}_{\sfL}+{\rm Ad}_{g_{\sfL}}\cL|_{z=-1}\,.
    \label{eq:A-Lax-hbiYB}
\end{align}
Since we use the same twist function (\ref{eq:twist-new}) with the symmetric coset case, 
we suppose the same ansatz for the Lax pair:
\begin{align}
    \cL=\left(U_{+}+z\,V_{+}\right)d\sigma^++\left(U_{-}+z^{-1}\,V_{-}\right)d\sigma^{-}\,.\label{eq:Lax-ansatz-bi}
\end{align}

\medskip

The solution (\ref{eq:A-bsol-bi}) leads to 
\begin{align}
    A|_{z=1}=2\eta_{\sfR}\,R_{\sfR}(\partial_{z}A|_{z=1})\,,\qquad 
      A|_{z=-1}=-2\eta_{\sfL}\,R_{\sfL}(\partial_{z}A|_{z=-1})\,.
      \label{eq:A-bsol-bi-re}
\end{align}
By using (\ref{eq:A-Lax-hbiYB}), (\ref{eq:Lax-ansatz-bi}) and (\ref{eq:A-bsol-bi-re}), we obtain
\begin{align}
    g^{-1}_{\sfR}\partial_{\pm}g_{\sfR}&=U_{\pm}+(1\mp2\eta_{\sfR}R_{\sfR,g_{\sfR}})(V_{\pm})\,,\\
    g^{-1}_{\sfL}\partial_{\pm}g_{\sfL}&=U_{\pm}-(1\mp2\eta_{\sfL}R_{\sfL,g_{\sfL}})(V_{\pm})\,.   
\end{align}
By solving these equations and removing $U_{\pm}$ from the Lax pair, we obtain the following expression: 
\begin{align} 
    \cL_{\pm}&=  g^{-1}_{\sfR}\partial_{\pm}g_{\sfR}-(1\mp\eta_{\sfR}R_{\sfR,g_{\sfR}})(V_{\pm})+z^{\pm1}V_{\pm} \notag 
    \\
    &= g^{-1}_{\sfL}\partial_{\pm}g_{\sfL}+(1\mp\eta_{\sfL}R_{\sfL,g_{\sfL}})(V_{\pm})+z^{\pm1}V_{\pm}\,, 
    \label{eq:Lax-biYB-L} 
\end{align}
where $V_{\pm}$ contains both $g_{\sfR}$ and $g_{\sfL}$ like 
\begin{align}    
    V_{\pm}&=\frac{1}{1\mp\eta_{\sfR}R_{\sfR,g_{\sfR}}\mp\eta_{\sfL}R_{\sfL,g_{\sfL}}}\left(  \frac{g^{-1}_{\sfR}\partial_{\pm}g_{\sfR}-g^{-1}_{\sfL}\partial_{\pm}g_{\sfL}}{2}\right)\,.
\end{align}

\subsubsection*{Deformed action}

Now, we can obtain the action of the homogeneous bi-YB deformed sigma model.
By using the expression of the Lax pair 
(\ref{eq:Lax-biYB-L})\,, 
the residues of $\varphi_c\,\cL$ at $z=\pm 1$ are evaluated as
\begin{align}
    \text{res}_{z=1}(\varphi_c\, \cL)&=4K(V_{+}d\sigma^{+}-V_{-}d\sigma^{-})\,,\\
        \text{res}_{z=-1}(\varphi_c\, \cL)&=-4K(V_{+}d\sigma^{+}-V_{-}d\sigma^{-})\,.
\end{align}
Then the 2D action becomes
\begin{align}
    S[g_{\sfR},g_{\sfL}]=K\int_{\Sigma}\langle g^{-1}_{\sfR}\partial_{+}g_{\sfR}-g^{-1}_{\sfL}\partial_{+}g_{\sfL},V_{-}\rangle d\sigma \wedge d\tau\,.
    \label{eq:bi-YB-action-lr}
\end{align}
This is an unusual form of the action of the homogeneous bi-YB deformed sigma model.

\medskip

In order to see the standard expression, let us use a complexified 2D gauge invariance 
$g_x\mapsto g_x h\,(h\in G^{\mathbb{C}})$\,.
Then, we can realize the following configuration:
\begin{align}
    g_{\sfR}=g\,,\qquad g_{\sfL}=1\,,
\end{align}
where $g \in G$\,.
With this gauge, the action (\ref{eq:bi-YB-action-lr}) reduces to
\begin{align}
    S[g]=\frac{K}{2}\int_{\Sigma}\left\langle g^{-1}\partial_{+}g,\frac{1}{1+\eta_{\sfR}R_{\sfR,g}+\eta_{\sfL}R_{\sfL}}g^{-1}\partial_{-}g\right\rangle d\sigma \wedge d\tau\,.
\end{align}
This is the standard expression of the homogeneous bi-YB deformed sigma model action. 
Then the Lax pair (\ref{eq:Lax-biYB-L}) is also simplified as 
\begin{align}
    \cL_{\pm}=\frac{1+z^{\pm1}\mp\eta_{\sfL}R_{\sfL}}{2}\left(\frac{1}{1\mp\eta_{\sfR}R_{\sfR,g}\mp\eta_{\sfL}R_{\sfL}}g^{-1}\partial_{\pm}g\right)\,.
\end{align}

\end{document}